\documentstyle[aps,prb,preprint,floats,eqsecnum,tighten,axodraw]{revtex}
\newcommand \be  {\begin{equation}}
\newcommand \ee  {\end{equation}}
\newcommand{\bq}{\begin{eqnarray}}
\newcommand{\eq}{\end{eqnarray}}

\newcommand{\bc}{\begin{center}}
\newcommand{\ec}{\end{center}}

\newcommand \w {\omega}
\newcommand \n {\eta}
\newcommand \no {\noindent}
\newcommand \eps {\epsilon}

\newcommand{\D}{\Delta}
\def\Dp{\Delta^{\prime}}

\def\cD{{\cal D}}

 \def\(({\left(}
 \def\)){\right)}

\def\[[{\left[}
\def\]]{\right]}

\def\a{\alpha}
\def\d{{\mbox d}}

\def\de{\delta}
\def\ap{\alpha^{\prime}}

\newcommand \s {\sigma}
\newcommand \f {\vec \varphi}
\input epsf

\begin{document}                

\begin{titlepage}
 
\title{{\bf Self Consistent Screening Approximation For Critical Dynamics}}
\author{Matteo Campellone, Jean-Philippe Bouchaud}

\date{\today}

\maketitle

\begin{center}
\vskip .2in
{ \it Service
de Physique de l'Etat Condens\'e, CEA-Saclay, Orme des Merisiers, 

91 191 Gif
s/ Yvette CEDEX, France \\
bouchau, matteo@amoco.saclay.cea.fr }
\vspace{2truecm}
\date{\today}
\begin{abstract} 
We generalise Bray's self-consistent screening approximation to 
describe the critical 
dynamics of the $\phi^4$ theory. In order to obtain the dynamical exponent 
$z$, we have to make an ansatz for the form of the scaling functions,
 which fortunately can be much constrained by general arguments.
 Numerical values of $z$ for $d=3$, and $n=1,...,10$ are obtained using
 two different ans\"atze, and differ by a very small amount. 
In particular, the value of $z \simeq 2.115$ obtained for the 3-d Ising model
 agrees well with recent Monte-Carlo simulations.

\end{abstract}

\end{center}

\hspace{.1in} PACS Numbers~: 02.50, 75.40G, 75.10H, 05.50
\end{titlepage}

\renewcommand{\baselinestretch}{2} 

\baselineskip=22pt 

\section{Introduction}

\no
Phase ordering kinetics, critical and low temperature dynamics of pure 
and random systems are the subject of active research \cite{Brayrev}. 
Of particular 
interest are the approximate methods to deal with non linear 
dynamical equations, which often amount to a self-consistent 
resummation of perturbation theory \cite{BCKM}. A much debated case is 
the `mode-coupling' approximation, used to describe liquids 
approaching their
frozen (glass) phase. Interestingly, this mode-coupling 
approximation for systems without disorder can alternatively be 
seen as the exact equations for  
an associated {\it disordered} model of the spin-glass type 
\cite{Krai,frhz,BCKM}. 
The simplest mode coupling approximation for the $\f^4$ theory is however
not very good. For example, it predicts for the static critical exponent
 $\eta$ the value $2 - \frac{d}{2} $ independently of the number 
$n$ of components of the field $\f$. Furthermore, the underlying disordered 
model is not stable \cite{BCKM}.

A better behaved resummation scheme is the ``Self-Consistent Screening 
Approximation" (SCSA) introduced by Bray in the context of the
 static
$\f^4$  theory \cite{bray1,bray2}, and used in other contexts 
\cite{MY,LD}. It amounts to resumming self
consistently all the diagrams appearing in the large-$n$
expansion, including those of order $1 \over n$. 
Again, this approximation becomes exact for a particular mean-field 
like spin-glass model \cite{BCKM}, which turns out to be well defined for all 
temperatures and thus ensures that the approximation is well behaved.

The aim of the present paper is to generalize the SCSA 
equations to describe the dynamics of the $\f^4$ theory {\it at the critical
point}, and to predict a value for the dynamical exponent $z$. 

In section \ref{scsa} we shall introduce the dynamical SCSA and the dynamical
equations in their general form. From section \ref{statics} and throughout
 the rest of the paper we assume that time-translation invariance (TTI) and the
fluctuation-dissipation theorem hold at least down to the critical point. 
Bray's equations will be recovered as the static limit of our
dynamical equations. The reliability of the SCSA is discussed
quantitatively in the $0$-dimensional static case.

In section \ref{dynamics} we study the equations right at the critical 
temperature where dynamical scaling is supposed to hold. 
The full solution of these coupled equations, involving {\it scaling functions} 
gives in principle 
the dynamical exponent $z$ within the SCSA approximation. Unfortunately, as 
is often the case \cite{KPZ}, these equations are very hard 
to solve, either analytically or even numerically. In section \ref{ansatz1} and
 \ref{ansatz2} we thus propose two different ans\"atze for the scaling 
functions, which are however much constrained by general requirements. 
The second ansatz leads to the exact $O(\eps^2)$ result in the $\eps=4-d$
RG expansion of Halperin, Hohenberg and Ma \cite{HHM}. The numerical value
 of the exponent $z$ only very weakly depends on the chosen ansatz,
 and turns out to be quite close to the best available Monte-Carlo estimate 
for the Ising model in $d=3$ \cite{Heuer}.

\section{The Self Consistent Screening Approximation}
\label{scsa}

Let us consider the coarse-grained Hamiltonian density
\be
{\cal H}[{\bf \f} (\vec{x})] = \frac{1}{2}(\nabla \f (\vec{x}))^2 + 
\frac{\mu}{2} \f^2 (\vec{x}) - 
\frac{g}{8} \f^4 (\vec{x}),
\label{h1}
\ee

\no
where ${ \f(\vec{x}) }$ is an $n$ component field and  
$\vec{x}$ is the $d$ dimensional space variable. With $\f^2 (\vec{x})$ 
and $\f^4
(\vec{x})$ we indicate respectively $|{\f(\vec{x}) }|^2$ and $(|{
\f(\vec{x}) }|^2 )^2$. The coupling constant
$g$ is negative and of order $n^{-1}$; $\mu$ is a (temperature dependent) 
mass term which 
vanishes at the mean-field transition point.
 
The partition function is 
\be
Z = \int \cD {\bf \f} e^{-\int d^{d}x  \ \frac{{\cal H}[{\bf \f}
 (\vec{x})]}{T}},
\ee

\no

In order to introduce the Self Consistent Screening Approximation one 
starts from
a large-$n$ expansion formalism. We re-write $Z$ with a gaussian 
transformation introducing an
auxiliary field $\s$
\be
Z = \int \cD \s \cD {\bf \f} e^{-\int d^{d}x \ \frac{ {\cal
H}[{\bf \f} (\vec{x}),\s(\vec{x})]}{T}} ,
\ee

\no
$H[{\bf \f} (\vec{x}),\s(\vec{x})]$ being now the Hamiltonian density of
two coupled fields
${\bf \f} (\vec{x})$ and $\s(\vec{x})$.

\be
H[\s,{\bf \f} ] = \frac{1}{2}(\nabla \f (\vec{x}))^2
+  \frac{\mu}{2}
 \f^2 (\vec{x}) + \frac{1}{2}\s^2 (\vec{x}) -
\frac{\sqrt{g}}{2}
\s (\vec{x}) \f^2 (\vec{x}) .
\ee

The SCSA amounts to consider the renormalization of the order 
$1/n$ diagrams in the Dyson expansion for the correlation functions of the 
two fields ${\bf \f} (\vec{x})$ and $\s(\vec{x})$.
Using this resummation scheme Bray \cite{bray1} obtained interesting
 results for the static
exponent $\n$  which describes the small momentum behaviour of
the correlation functions.  The static SCSA equations for
 $\langle {\bf \f} (\vec{x})
{\bf \f} (\vec{x}^{\prime}) \rangle$ (plain line) and
$\langle \s(\vec{x}) \s(\vec{x}^{\prime}) \rangle$ (``dashed" line) are reported
diagrammatically in figure 1.
The bare quantities are indicated respectively with a thinner plain
 line and with a dashed line.

\begin{figure}
\epsfbox{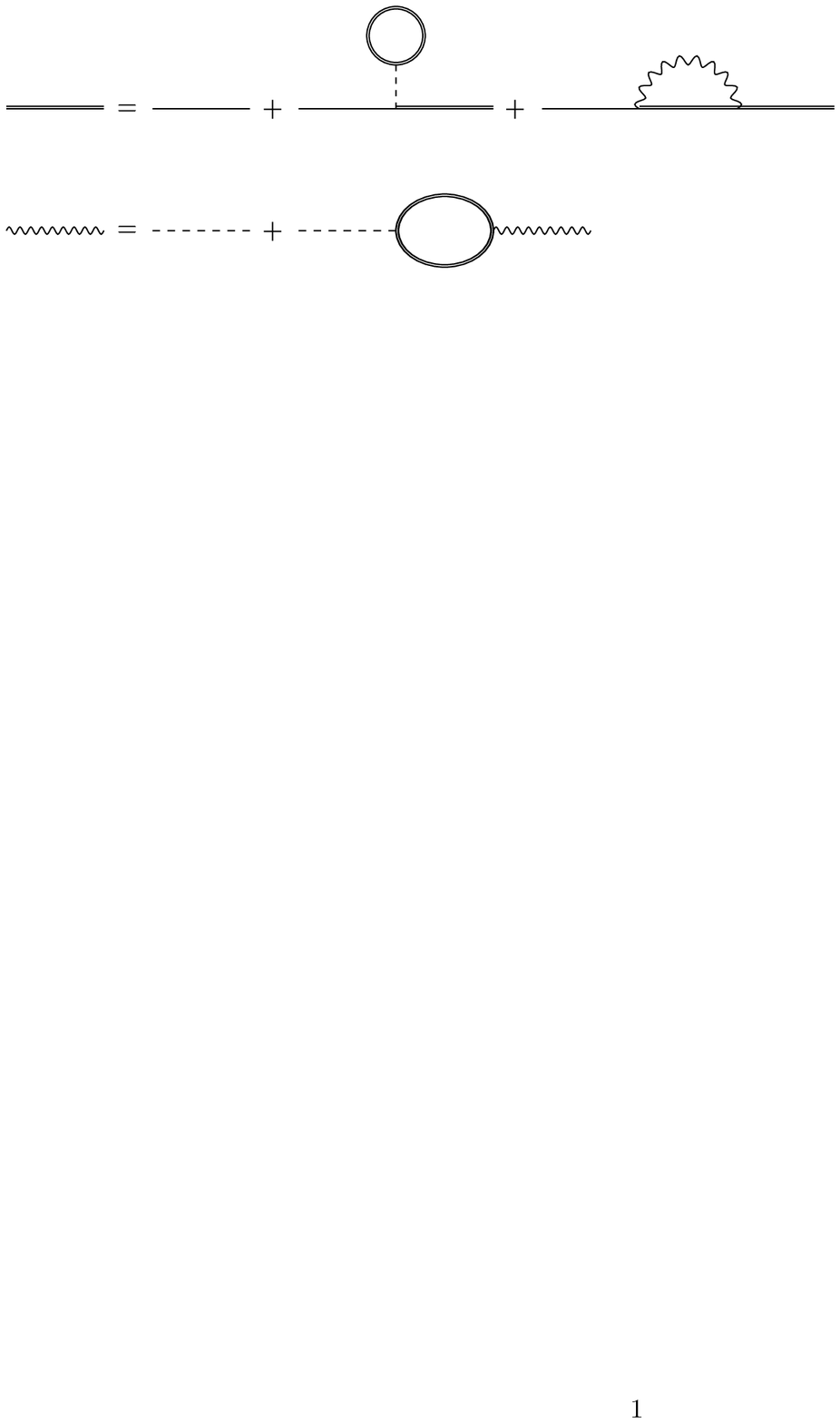}
\vspace{0.5cm}
\caption[]{}
\end{figure}

Our goal is to develop a dynamical generalization of this expansion for 
non-conserved Langevin dynamics, starting from the SCSA Hamiltonian. We thus
obtain the following equations of motion for ${\bf \f} (\vec{x},t)$ and
$\s(\vec{x},t)$:
\bq 
\dot{{\bf \f} }(\vec{x},t)& = &-(\nabla^2 + \mu)
 {\bf \f} (\vec{x},t) + \sqrt{g} {\bf \f} 
(\vec{x},t) \s (\vec{x},t) +
\eta_{\f} (\vec{x},t) 
\label{langf} \\
\dot{\s}(\vec{x},t)& = & -  \s (\vec{x},t) + \frac{\sqrt{g}}{2}
\f^2(\vec{x},t) + \eta_{\s}(\vec{x},t).
\label{langs}
\eq
\noindent
with two independent thermal noises $\eta_{\f}, \eta_{\s}$. 

Let us now consider the two-point functions
\bq
G_{\f}(\vec{x},\vec{x}^{\prime},t,t^{\prime}) &=& \left<
\frac{\partial{{\bf \f}(\vec{x},t)}} {\partial{\n(\vec{x}^{\prime},t
^{\prime})}}
\right> \\ C_{\f}(\vec{x},\vec{x}^{\prime},t,t^{\prime}) &=& < {\bf \f}
(\vec{x},t)
{\bf \f}(\vec{x}^{\prime},t^{\prime}) >,
\eq
and the corresponding functions for the field $\s$. The SCSA dynamical 
equations, which can be seen as a Mode-Coupling approximation on the set of
equations (2.5-6) (see figure 2) then read:

\begin{eqnarray} \nonumber
\Sigma_{\f}(t_1,t_2) & = & 
 n \frac{g}{2} \delta(t_1-t_2)  \int_{0}^{t_{1}} dt_{3} C_{\f}(t_{3} ,t_{3}) 
G_{\s}^{0} (t_{1},t_{3})  + 
   \\
&+& g [G_{\f}(t_{1},t_{2})C_{\sigma}(t_{1},t_{2})   
+ 
G_{\s}(t_{1},t_{2})C_{\f}(t_{1},t_{2})] \label{2times1}
\eq
\be
\Sigma_{\s}(t_1,t_2) =   n g  G_{\f}(t_{1},t_{2})
C_{\f}(t_{1},t_{2}) 
\ee
\bq
D_{\f}(t_1,t_2) &=&  2T \de (t_1-t_2) + 
g C_{ \f}(t_1,t_2) C_{ \s}(t_1,t_2) 
   \\ 
D_{\s}(t_1,t_2) &=& 
2T \de (t_1-t_2) + n \frac{g}{2} C_{\f}^2 (t_1,t_2),
\label{2times}
\eq
where we have dropped the space coordinates $\vec x$ for clarity,
 and introduced the self-energies $\Sigma$, defined as:
\be
G(t,t^{\prime}) = G^0(t,t^{\prime}) + \int_{0}^{t} dt_{1} \int_{t'}^{t_1}
dt_2 G^{0}(t,t_{1}) \Sigma(t_{1},t_{2}) G (t_{2},t^{\prime}),
\ee
(the label $^0$ refers to the bare quantity), and the `renormalized noises'
$D$, defined as:
\be
C(t,t^{\prime}) = \int_{0}^{t}dt_{1} \int_{0}^{t^{\prime}}dt_{2}
 G(t,t_{1})D(t_{1},t_{2})G(t^{\prime},t_{2}).
\label{cdef}
\ee

\begin{figure}
\epsfbox{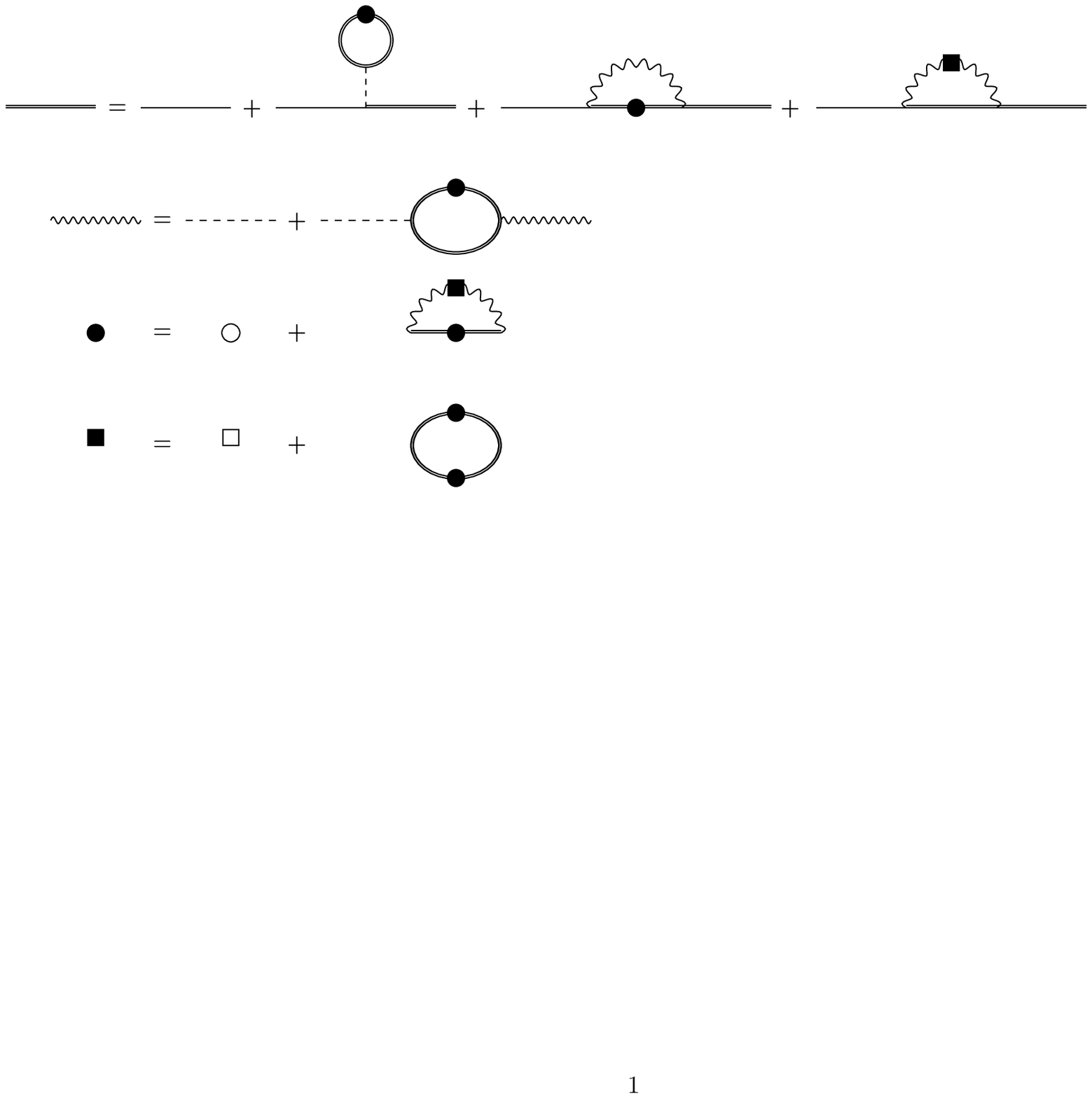}
\vspace{0cm}
\caption[]{}
\end{figure}

We shall limit ourselves to consider the above equations
 in a regime of stationary dynamics. 
That is to say that we will make use of the assumption of
 time translational invariance (only differences of times matter), 
which allows one to show that
 the fluctuation dissipation theorem (FDT) is valid, i.e:
\be
\theta(t-t^{\prime}) \frac{\partial C(t-t^{\prime})}{\partial{t^{\prime}}} =
T G(t-t^{\prime}).
\ee
Extensions of these methods to non stationary low temperature
 regime, where this theorem is violated \cite{CK}, will
 be subject of further work. In the following, we shall set
 the energy scales by choosing $T=1$, and vary the mass term $\mu$ to
 reach the critical point.

\section{Static Limit}
\label{statics}

With these assumptions equations (\ref{2times}) reduce to only
two coupled independent equations which have the simplest form in Fourier
space

\bq \nonumber
\Sigma_{\f}(k,\w) &=& g \int \left[ C_{\s}(k-
k^{\prime},
\w - \w^{\prime})  G_{\f}(k^{\prime},\w^{\prime}) + 
C_{\f}(k-k^{\prime},\w - \w^{\prime})
 G_{\s}(k^{\prime},\w^{\prime}) \right] Dk'D\w'  
\\
 && +  \frac{ n g}{2} G_{\s}^{0}(k=0,\w=0) 
\int C_{\f}(k^{\prime},\w^{\prime})   Dk'D\w' 
\label{sigf}  \\  
\Sigma_{\s}(k,\w) &=& {n g} \int  C_{\f}(k-
k^{\prime},\w - \w^{\prime})
 G_{\f}(k^{\prime},\w^{\prime})
 Dk'D\w' .   
\label{sigs}
\eq 
where $Dk' \equiv \frac{d^d k'}{(2\pi)^d}$ and $D\w' \equiv \frac{d\w'}{2\pi}$.

Using the fact that $C(k,t=0) \equiv {\cal C}(k)$ is equal to $G(k,\omega=0)$
(from FDT and the Kramers-Kronig (KK) relations), and using again
 the KK relations, it is easy to check that for $\w=0$ one recovers exactly
 the static  
SCSA equations \cite{bray1}, namely

\bq
{\cal C}_{\f}(k) & = & \frac{1}{\mu + k^2 - g \int Dk'
 {\cal C}_{\f}(k-k'){\cal C}_{\s}(k') -\frac{g n}{2} 
\int Dk' {\cal C}_{\f}(k')} \nonumber \\ {\cal C}_{\s}(k)
 & = & \frac{1}{1 - \frac{g}{2}  n
\int Dk'{\cal C}_{\f}(k-k'){\cal C}_{\f}(k')},
\label{cstatiche}
\eq

\noindent

In order to test the validity of this approximation, it is 
interesting to consider the case of zero spatial dimensions \cite{bray2}. Let us
set $n=1$ which is a bad case for the SCSA which should become more
accurate the larger $n$ is. We will compare equations (\ref{cstatiche}) with
the exact static
 correlation function which in zero dimension can be calculated
 analytically and is
\be
  {\cal C}_{exact} = -{1\over { \mu}} + {\mu \over g} - 
   {{\mu \,{\rm K}_{-\frac{3}{4}}(
        {{{\mu^2} }\over {4\,g}})}\over 
     {2\,g\,{\rm K}_{1\over 4}(
        {{- \,{\mu^2} }\over {4\,g}})}} - 
   {{\mu \,{\rm K}_{5\over 4}(
        {{{\mu^2} }\over {4\,g}})}\over 
     {2\,g\,{\rm K}_{1\over 4}(
        {{- \,{\mu^2} }\over {4\,g}})}},
 \ee

\no
where $K_{n}(a)$ is the modified Bessel function of the second kind.
Equations (\ref{cstatiche}) give for ${\cal C}_{\f}$:

\be
{\cal C}_{scsa}  =  \frac{1}{\left(\mu - 
 n \frac{g}{2} {\cal C}_{scsa} - g \frac{{\cal C}_{scsa}}
{ \left( 1 - \frac{g}{2}  n
{\cal C}_{scsa}^2 \right)}\right)} .
\ee

From plotting the relative
 difference of the two correlation functions versus the coupling (see
figure 3)
 we can see that SCSA is quite close to the exact theory.
In particular, the asymptotic behaviour in the $|g| \rightarrow \infty$ 
limit of the two 
functions is 

\be
\lim_{|g| \rightarrow \infty}  \sqrt{|g|}
{\cal C}_{scsa} = 2(\sqrt{2} - 1)  {\mbox{       and       }}  
\lim_{|g| \rightarrow \infty}   \sqrt{|g|} C_{exact} =  
{{2\,{\sqrt{2}}\,{\Gamma}({3\over 4})}\over 
    {{\Gamma}({1\over 4})}}.
\ee

For all $g$, the relative difference is actually bounded by:

\be
\frac{\left| {\cal C}_{exact} - {\cal C}_{scsa} \right|}
{{\cal C}_{exact}}  
<  1 - {{{\sqrt{-1 + {\sqrt{2}}}}\,{\rm \Gamma}({1\over 4})}\over 
     {2\,{\rm \Gamma}({3\over 4})}}
=  0.0479...
\ee

\no
We can also compare the small$-g$ 
expansions of the two theories which give

\bq
{\cal C}_{exact} &=& \frac{1}{ \mu}(1 + \frac{3}{2 \mu^2} g +
 \frac{21}{4 \mu^4} g^2)  \\ 
{\cal C}_{scsa} &=& \frac{1}{ \mu}(1 + \frac{3}{2 \mu^2} g +
 \frac{5}
{\mu^4} g^2)
\eq
 showing explicitly how the two theories 
differ already at order $g^{2}$. 
The self consistent nature of the approximation however keeps the SCSA in good
agreement with the exact theory even for large values of the coupling constant
 as remarked before.

It is instructive, in passing, to compare the SCSA with the simple
 Hartree ($n = \infty$) resummation scheme, which is also the 
Gaussian variational result. One defines $F_{H} = \min{\{F\} }$ where

\be
F = F_{0} + <H-H_{0}>,
\ee

with
\be
F_{0} = - \ln \int  \cD {\bf \f} e^{-\frac{\tilde{\mu}\f^2}{2}}  =
- \ln{(\frac{2 \pi}{\tilde{\mu}})}
\ee
\be
<H_{0}> = \frac{1}{2 }
\ee
\be
<H> =  \int  \cD {\bf \f} e^{-\frac{\tilde{\mu}\f^2}{2}}  \left(
\frac{\mu}{2}\f^{2} - \frac{g}{8}\f^{4} \right)  = \left(
\frac{\mu}{2\tilde{\mu}} - \frac{3g}{8 \tilde{\mu}^{2}} \right)
\ee
Minimising $F$ with respect to $\tilde{\mu}$ we find 
\be
\mu_{H} = \frac{\mu + \sqrt{\mu^2 - 6g}}{2},  
\ee
and consequently

\be
{\cal C}_{H} =  <\f^2>_{\mu_{H}} = \frac{2}{\mu + \sqrt{\mu^2 - 6g}}.
\ee

As can be seen from figure 3, the SCSA turns out to be fairly better
 than the Hartree variational approach (at least in this particular case 
of $n=1$ and $d=0$).

\begin{figure}
\epsfbox{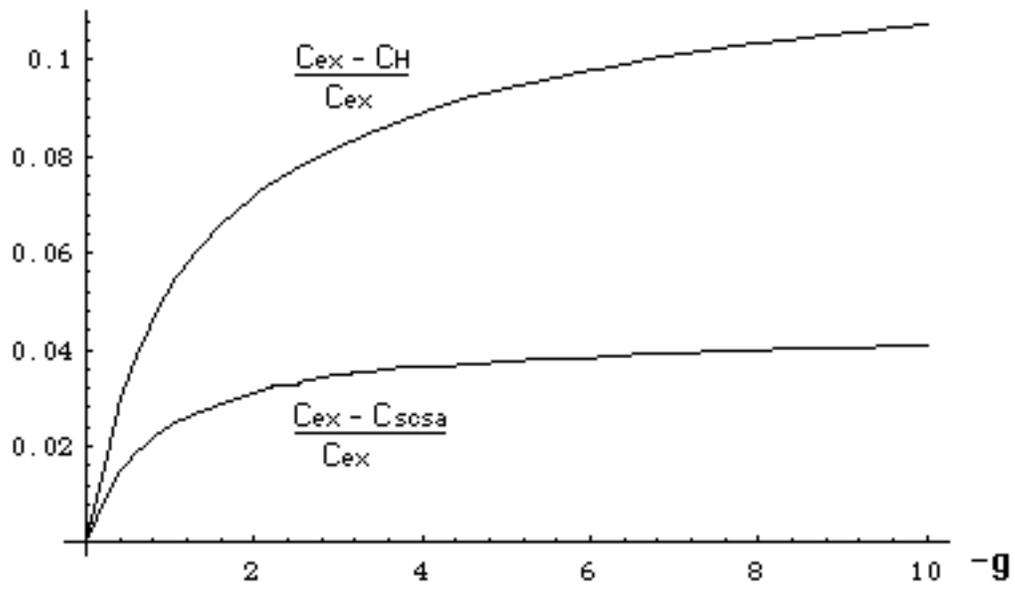}
\vspace{0.5cm}
\caption[]{Relative difference between the exact result and the Hartree 
(${\cal C}_H$) and the SCSA (${\cal C}_{scsa}$) approximations,
 in the case $n=1,d=0$.}
\end{figure}

\section{Critical Dynamics}
\label{dynamics}

We shall now work right at the critical point $\mu_{c}$ such that 
the renormalised mass vanishes (therefore eliminating the `tadpole' 
contribution in Eq. \ref{2times1}). We shall search for solutions under
 the general dynamic scaling form (valid in the small-$k$ and 
small-$\w$ limit):

\bq
G_{\f}(k,\w) &=& \frac{1}{k^{\D}} n_{\f}(\frac{\w}{k^{z}}) 
\hspace{2truecm} G_{\s}(k,\w) =
\frac{1}{k^{\D^{\prime}}} n_{\s}(\frac{\w}{k^{z}})
 \nonumber
\\ C_{\f}(k,\w) &=& \frac{2}{ \w k^{\D}} Im \left[ 
n_{\f}(\frac{\w}{k^{z}}) \right ]
\hspace{.7truecm} C_{\s}(k,\w) = \frac{2}{ \w k^{\D^{\prime}}} 
Im \left[
n_{\s}(\frac{\w}{k^{z}})
\right ]. 
 \label{generalscaling}
\eq  where we have defined $\D = 2- \n$, and used FDT.
\no
Setting first $\w=0$, one finds by matching the momentum dependence of the 
left and right hand sides of (\ref{sigf}-\ref{sigs}) that: 
\be
\D^{\prime} = d - 2 \D = d - 4 + 2 \n.
\ee
Note that in mean field, $z=2$, $\D =2$, $\eta=0$ and $\D'=0$.
Identification of the prefactors yields: 
\be
n_{\s}(0) n_{\f}^2 (0) = - \frac{2}{ f(\eta,d) n g }
\label{relstat}
\ee
where
\be
f(\eta,d) = \frac{1}{(4 \pi)^{d/2}} \frac{\Gamma[\D - \frac{d}{2}]
 \Gamma[\frac{d-\D}{2}]^2 }{\Gamma[d-\D] \Gamma[\frac{\D}{2}]^2} .
\label{fbray}
\ee
and an extra equation fixing $\eta$ as a function of $d$ and $n$, which 
we do not write explicitely \cite{bray1}. 

Now let us consider the other case where $k=0$ and $\w > 0$ (but small).
Taking the imaginary part of (\ref{sigf}-\ref{sigs}), one obtains:

\bq
Im \left[  \Sigma_{\f}(0,\w) \right]  &=& \frac{S \w }{n n_{\f} (0)}
 \int q^{\Delta -1} dq ds
 \frac{Im\left[ f_{\f} (\frac{(\w -s)}{q^{z}}) \right] 
 Im\left[ f_{\s} (\frac{s}{q^{z}}) \right] }{s (\w -s)}
\label{fk=0} \\
Im \left[ \Sigma_{\s}(0,\w)\right]   &=& \frac{S}{n_{\s}(0)} \int
  q^{\Delta ^{\prime}  -1} dq  d s  \frac{Im\left[ f_{\f}
 (\frac{(\w -s)}{q^{z}}) \right]  Im\left[ f_{\f} 
(\frac{(s)}{q^{z}}) \right] }{s},
 \label{sk=0}
\eq

\no
where $f_{\f,\s}(x) = n_{\f,\s}(x)/n_{\f,\s}(0)$. We also defined

\be
S = \frac{2 n g \Omega_{d}}{(2 \pi)^{(d+1)}}  n_{\f}^2(0) n_{\s}(0)
\equiv -\frac{4 \Omega_{d}}{f(\eta,d)(2 \pi)^{(d+1)}} \label{eq1}
\ee
In general the scaling functions can be written 
\bq
Im[f_{\f}(x)] & \doteq & A \tilde{f_{\f}} (a x)    \\ \nonumber
Im[f_{\s}(x)] & \doteq & A'\tilde{f_{\s}} (a^{\prime} x),
\eq
\no
with by convention $\lim_{u \to \infty}u^{\D/z} \tilde{f_{\f}}(u) =1$ and 
$\lim_{u \to \infty} u^{\D'/z} \tilde{f_{\s}}(u)=1$. This asymptotic behaviour
is required for the $k \to 0$ limit to be well defined, if 
(\ref{generalscaling}) is correct. Furthermore, the small-$\w$
 behaviour of
the imaginary part of the response function is expected to be regular 
for $k$ finite, and hence $\tilde{f}(u) \propto u$ for $u \to 0$. 
$A,A'$ are coefficients setting the scale of the imaginary part of 
the response function while $a,a'$ are coefficients 
setting the frequency scales. 
Using the fact that the imaginary and real part of the response function are
 power-laws at large frequencies, which imply that their ratio is 
$\tan\left( \frac{\pi \D}{2z} \right)$ 
(resp. $\tan\left( \frac{\pi \D'}{2z} \right)$), one finds that:

\bq \nonumber
\frac{a^{\D/z} }{A} \sin^{2}
 \left( \frac{\pi \D}{2z} \right) &=& 
\frac{ S }{nz} \int_{0}^{\infty} 
\frac{dx}{x^{1 + \D/z}} \int_{-\infty}^{\infty} \frac{du}{u(1-u)}
 Im\left[ f_{\f} (x (1-u)) \right]  Im\left[ f_{\s}(x u) \right] \\
\frac{\a'^{\Dp/z} }{A'}
 \sin^{2} \left( \frac{\pi \Dp}{2z} \right) &=& \frac{ S }{z} 
\int_{0}^{\infty} \frac{dx}{x^{1 + \Dp/z}}
 \int_{-\infty}^{\infty} \frac{du}{u} Im\left[ f_{\f} (x (1-u)) \right] 
 Im\left[ f_{\f} (x u) \right] 
\label{scsagens}
\eq

It is easy to show that these equations actually only depend on the value of
the {\it ratio} of frequency scales $y=\frac{a'}{a}$. The coefficient $A$ 
can be fixed using the KK relation, since the involved integral converges, 
which means that the small-$k$
 behaviour of the real part of the correlation function is
fully determined by the imaginary part in the scaling region $\omega, k \to 0$.
Hence:
\be
1 = \frac{A}{\pi} \int_{-\infty}^{\infty} dx \frac{\tilde{f_{\f}} (x )}{x}.
\label{kk} 
\ee
The corresponding integral for $\tilde{f_{\s}}$ does not converge for 
large $x$, meaning that the non-scaling region is needed to saturate the 
sum-rule. Hence, we must use another relation to fix $A'$, which we
 choose to be the small-$\w$ expansion of Eq. (\ref{sk=0}). 

Thus, {\it if} the functions $\tilde{f_{\f}},\tilde{f_{\s}}$ were
 known, we would have four equations to fix four constants: $A,A',y$,
 and, of course, the dynamical exponent $z$, in
 terms of $d$ and $n$. $\tilde{f_{\f}},\tilde{f_{\s}}$
are in principle also fixed by the full equations for
 arbitrary $\frac{\omega}{k^z}$. However,
 as in other similar cases \cite{KPZ}, these equations are very
 hard to solve, either analytically or numerically. We will
 thus propose ans\"atze for these functions, which have to satisfy the
 above general requirements. Note that once $A,A',a,a'$ have been pulled
 out, the only
freedom is in the {\it shape} of these functions. We shall thus work
 with two such ans\"atze, which will turn out to give very similar answers 
for $z$. This
was also the case in the context of the KPZ equation \cite{KPZ}.

\section{Ansatz 1}
\label{ansatz1}

The simplest ansatz one can think of, which generalizes the mean field
shape:
\be
\tilde{f_{\f}}(x) = \frac{x}{(1+x^2)} 
\label{fmf}
\ee 
reads:
\bq
\tilde{f_{\f}}(x) &=&  \frac{x}{\left(1+x^2 \right)^{\a}} \\
\tilde{f_{\s}}(x) &=&  \frac{x}{\left(1+x^2\right)^{\ap}} ,
\eq
where we have set 
\bq
\a &\doteq& \frac{\D + z}{2z} \\
\ap &\doteq& \frac{\Dp + z}{2z}.
\label{alphas}
\eq
(Note that $\a=1$ in mean field). 
These functions have indeed the correct asymptotic behaviours; they go linearly
to zero for small values of the argument and behave as power laws
($\tilde{f_{\f}}(x) \simeq x^{-\frac{\D}{z}} $ and $
\tilde{f_{\s}}(x) \simeq x^{-\frac{\Dp}{z}}$) in the large-$x$ limit.

We can now use (\ref{kk}) to determine $A$ 
\be
A = \sqrt{\pi} \frac{\Gamma[\a]}{\Gamma\left[\a - \frac{1}{2} \right]}.
\ee

The small-$\w$ expansion of $Im \Sigma_{\s}(k,\w) $ can be matched with that of
 the right hand side of Eq.(\ref{sk=0}) leading to
the following equation
\be
y = -\frac{2 A^2}{A' f(\eta,d) (2 \pi)^{d+1}}
 \int_{-\infty}^{\infty} d^d q \frac{1}{|q|^{\D} |1-q|^{\D + z}}
 \int_{-\infty}^{\infty} dt
\frac{1}{(1+t^2)^{\a} \left(1+ (\frac{|q|^{z} t}{|1-q|^{z}} )^2 \right)^{\a}}
\label{ggg}
\ee

After some algebraic manipulations we obtain for the last three equations:

\bq \nonumber
\sin^{2} \left( \frac{\pi \D}{2z} \right)  &=&- \frac{A^2 A'y S}{2 nz}  
B\[[1 -\frac{\D}{2z}, \frac{d}{2z}\]] \\& &
 \int_{-\infty}^{\infty} \frac{du}{ |u|^{2-\frac{\D}{z}}}  F\[[\ap,1
-\frac{\D}{2z},\a + \ap,1- y^2 \frac{(1-u)^2}{u^2}\]] \label{scsa11}
\eq
\bq \nonumber
  \sin^{2} \left( \frac{\pi \Dp}{2z} \right) &=&- \frac{A^2 A^{\prime}S}{2z}
y^{-\frac{\Dp}{z}} B \[[1
-\frac{\Dp}{2z}, \frac{d}{2z}\]]\\
& & \int_{-\infty}^{\infty}
 \frac{u du}{|u|^{2 -\frac{\Dp}{z}}} F\[[\a,1 -\frac{\Dp}{2z},2
\a,\frac{2u-1}{u^2}\]] 
\label{scsa12}
\eq
\bq \nonumber
 y &=& \pi  \frac{A^2 S}{z A' \Omega_{d}}
 B\[[\frac{1}{2},2 \a -\frac{1}{2}\]] \\
& & \int_{0}^{\infty}
dq q^{d-2-\D}
\int_{|1-q|^{2z}}^{|1+q|^{2z}} \frac{dx}{x^{\frac{\D +3z- 2}{2z}}}
F\[[\a,\frac{1}{2},2\a,1-\frac{q^{2z}}{x}\]],
\label{scsa13}
\eq 
where $B[a,b]$ and $F[a,b,c,x]$ are the Euler Beta and Hypergeometric
functions and where the last equation (\ref{scsa13}) was written for the 
special case $d=3$ which we shall consider below.
We can solve analytically Eqs. (\ref{ggg},\ref{scsa11},\ref{scsa12}) 
at order $\eps^2$ to compare 
with the exact RG treatment of \cite{HHM}. At lowest order we obtain:
 \bq
 c &=& \frac{8 \ln{2}}{ \pi}   
\frac{ \arctan \sqrt{\frac{1 - y^2}
{y^2}}}{\sqrt{1 - y^2}}-1  \\
A' &=& - \frac{\pi \eps}{4}  \\
y &=& \frac{4 \ln 2 }{\pi}
\label{oeps2}
\eq
where we have defined, following \cite{HHM},
\be
z = 2 + c \eta .
\label{formz}
\ee
The order $O(\eps^2)$ RG result reads, $c =6\ln{\frac{4}{3}} -1 =0.7261$.
 The form (\ref{formz}) means that to lowest order $z$ depends on $n$ 
only through the static exponent $\n$.  On the other hand, 
Eqs. (\ref{oeps2}) give 
\be
c = 0.8376,
\ee
in slight disagreement with the exact result. 
This comes from the fact that while our ansatz for $\tilde f_{\f}$ 
is exact in the limit $\epsilon \to 0$, 
the corresponding ansatz for $\tilde f_{\s}$ is already wrong at lowest 
order since it does not satisfy Eq. (\ref{sk=0}).
 In our second ansatz, we thus keep
the same shape for $\tilde f_{\f}$, but choose 
for $\tilde f_{\s}$ a form which is
exact when $\epsilon \to 0$.

\section{Ansatz 2}
\label{ansatz2}

Knowing the mean field form for $f_{\f}(x)$ we can, at lowest
order in $\epsilon$, write for $Im[f_{\s}(x)]$

\be
Im f_{\s}(x) = 2^{d-4} \frac{ f(\eta,d)}{\pi^{d/2}} 
\frac{(2\pi)^d}{\Gamma[2-\frac{d}{2}]} Im \[[ \frac{1}{\xi(x)} \]] 
\label{gsgen}
\ee
where
\be
\xi(x) = 1 - \frac{\eps}{2} \int_{0}^{1} dt \log\[[ 1-t^2 - 2 i x (1-t) \]].
\ee
 It is then straightforward to generalize $Im[f_{\s}(x)]$ to general dimensions
as:
\be
\tilde {f_{\s}} \propto  Im \[[ (2 - i  x )^{1 -
\frac{\Dp}{z}} - (1 - i  x )^{1 - \frac{\Dp}{z}} \]]
\label{fsfin}
\ee
with a prefactor ensuring that the coefficient of $x^{-\frac{\Dp}{z}}$ 
for large $x$ is unity.  Eq. (5.8) is now 
replaced by: \be
\sin^2 \((\frac{\pi \D}{2 z} \)) = \frac{A^2 A' S b}{n z} \int_{0}^{\infty}
\frac{\d r}{r^{\frac{\D}{z}}} \int_{-\infty}^{\infty} \d u 
\frac{Im \[[ (2 - i \frac{\pi}{2
\ln{2}}  (y r u) )^{1 - \frac{\Dp}{z}} -
(1 - i \frac{\pi}{2 \ln{2}} 
(y r u) )^{1 - \frac{\Dp}{z}} \]]}{u \[[1 + r^2 (1-u)^2 \]]^{\a}}.
\label{scsa21}
\ee
where now $b$ is given by:  
\be
b = \frac{2 \ln{2}}{\pi  \((2^{ - \frac{\Dp}{z}} -1\)) 
\((\frac{\Dp}{z} -1 \)) }.
\ee

We finally obtain a set of equations for $z$ of 
the same kind as (\ref{scsa11}-\ref{scsa13}) but which now exact up to
 $O(\epsilon^2)$, as we have checked directly.

\section{Numerical Results}

We solved  numerically both sets of equations in $d= 3$ for $n=1,...,10$.
We used the values of $\eta(d=3,n)$ that can be derived from the
 formula reported in
\cite{bray1}.
The values obtained for $z$ are reported in the following table.

\medskip
\begin{equation}
\begin{array}{|c|c|c|} \hline       
\par
n & z \mbox{ (ansatz 1 )} & z \mbox{ (ansatz 2 )}\\ \hline   
1&  2.119 & 2.113   \\ \hline  
2 & 2.071 & 2.069   \\ \hline   
3 & 2.050& 2.049     \\ \hline 
4 & 2.038 & 2.038   \\ \hline  
5 & 2.031& 2.031     \\ \hline 
6 & 2.0258& 2.0258   \\ \hline 
7 & 2.0223& 2.0222   \\ \hline 
8 & 2.0196& 2.0195   \\ \hline 
9 & 2.0174& 2.0174   \\ \hline 
10 & 2.0157& 2.0157   \\ \hline 
\end{array}
\nonumber
\end{equation}
\medskip

As it was hoped, the results are fairly independent from 
the ansatz used, which is more and more true for large $n$. The result 
for $n=1$ is rather close to the best Monte-Carlo estimate 
of ref. \cite{Heuer}, which gives $z=2.09 \pm 0.02$. Let us note 
however
that the SCSA overestimates significantly $\eta$ in $d=3$.

In figure (4) we compare the two different choices for the 
scaling function $f_{\s}(x)$ with their relative values of the
 parameters $y$, $A'$ and $z$, and in the case 
$n=1,d=3$. We notice that the constraints for small $x$ and
 large $x$ restrict very much the freedom on the shape of this function. 

\begin{figure}
\epsfbox{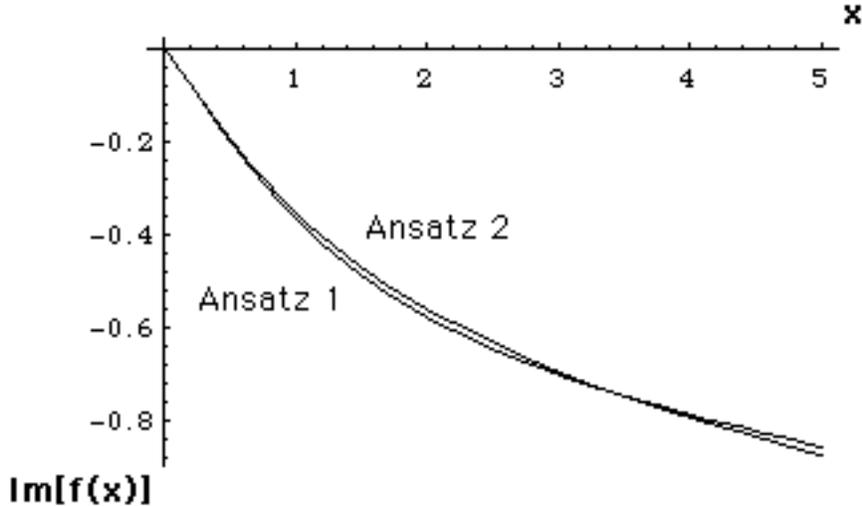}
\vspace{0.5cm}
\caption[]{The two ans\"atze for the functions $f_{\s}(x)$, $n=1$}
\end{figure}

Finally, a linear regression of our results for $n=1-10$
 gives $z \simeq 2 + c  \eta$ with $c=0.64$, which is lower
 that the $O(\eps^2)$ result, but larger that
the exact result for $d=3$, $n \to \infty$, i.e. $c=\frac{1}{2}$ \cite{HHM}. 

\section{Conclusions}

The aim of this paper was extend the static Self-Consistent screening
 approximation to dynamics, in particular to calculate the properties of the
critical dynamics of the $\phi^4$ model. Although the resulting
 equations cannot be fully solved, a much constrained ansatz leads to a value 
of the exponent $z$ in rather good agreement with Monte-Carlo data.

Our work was originally inspired by glassy dynamics: the
SCSA equations actually describe in exactly the dynamics of some mean-field 
spin glass like models. It would be interesting to study these equations in 
the low temperature phase, where dynamics becomes non stationary (aging)
 and FDT is lost. For $\phi^4$ models, this corresponds to a coarsening
 regime \cite{Brayrev}. It would be
 interesting to know whether the SCSA equations describe properly this
regime, and can compete with other approximation schemes
\cite{Brayrev,NB}.

\vskip 1cm 
{\it Acknowledgments} It is a pleasure for us to thank A.~Barrat, 
A.~Bray, L.F.~Cugliandolo, J.~Kurchan,
E.~Maglione, M.~M\'ezard, R.~Monasson, G.~Parisi and P.~Ranieri for
very instructive discussions.


\bibliographystyle{IEEE}

\newpage

Caption for figure 1:

Diagrammatic equations for the correlation functions 
$\langle {\bf \f} (\vec{x})
{\bf \f} (\vec{x}^{\prime}) \rangle$ (plain line) and
$\langle \s(\vec{x}) \s(\vec{x}^{\prime}) \rangle$ (dashed line).

Caption for figure 2:

Diagrammatic representation of the dynamical SCSA equations, 
where the
full circle stands for the renormalization of $D_{\f}$ while the full square for
the renormalization of
$D_{\s}$. The empty circle and empty square stand for the 
non-renormalized noises.

Caption for figure 3:

Relative difference between the exact result and the Hartree 
(${\cal C}_H$) and the SCSA (${\cal C}_{scsa}$) approximations,
 in the case $n=1,d=0$.

Caption for figure 4:

The two ans\"atze for the functions $f_{\s}(x)$, $n=1$

\end{document}